# Structural stability and polarization analysis of rhombohedral phases of HfO$_2$

Wenbin Ouyang[1], Fanhao Jia[1], Silvia Picozzi[2] and Wei Ren[1,*]

[1]*Physics Department, International Center for Quantum and Molecular Structures, Materials Genome Institute, State Key Laboratory of Advanced Special Steel, Shanghai Key Laboratory of High Temperature Superconductors, Shanghai University, Shanghai 200444, China*

[2] *Consiglio Nazionale delle Ricerche CNR-SPIN, c/o Univ. "G. D'Annunzio", 66100 Chieti, Italy*

\* renwei@shu.edu.cn

A comparative theoretical study is presented for the rhombohedral R3 and R3m phases of HfO$_2$, i.e. two possible forms of heavily Zr-doped ferroelectric thin films of Hafnia, recently found in experiments. Their structural stability and polarization under in-plane compressive strain are comprehensively investigated. We discovered that there is a phase transition from R3 to R3m phase under biaxial compressive strain. Both the direction and amplitude of their polarization can be tuned by strain. By performing a symmetry mode analysis, we are able to elucidate the improper nature of ferroelectricity. These results may help to shed light on the understanding of ferroelectricity in hafnia thin films.

## I. INTRODUCTION

Ferroelectric thin films that are compatible with modern silicon-based information technologies are highly desired for non-volatile memory and advanced energy devices [1-4]. $HfO_2$ has long been studied in the microelectronics industry, due to its high dielectric constant that could improve the leakage current problem in the gate layer of $SiO_2$ films [5,6]. In the past decade, the ferroelectricity in orthorhombic $HfO_2$ thin films has been exhaustively investigated by both experimental and theoretical studies, as ferroelectric Hafnia is considered to be the most promising ferroelectric gate dielectric material, due to its compatibility with the contemporary metal-oxide-semiconductor-based devices [7,8].

The orthorhombic ferroelectric phase of $HfO_2$ films with a space group of $Pbc2_1$ was firstly observed in 2011 by Böscke *et al.* using X-ray diffraction measurements [9], and was later confirmed by scanning transmission electron microscopy [10]. In 2018, Wei *et al.* experimentally observed the rhombohedral (R) phase of $Hf_{0.5}Zr_{0.5}O_2$ (HZO) films with (111) direction grown on the (001)-oriented $La_{0.7}Sr_{0.3}MnO_3/SrTiO_3$ substrate [11], showing an increased out-of-plane polarization upon decreasing the film thickness. First-principles simulations proposed the predicted R3m phase under epitaxial compression, and even the R3 polymorph predicted in the Hf-rich limit may be approximate representations of the rhombohedral structure in the actual samples [11]. More recently, Bégon-Lours *et al.* analyzed the arrangement pattern of Hf atoms in HZO films epitaxially grown on GaN(0001)/Si(111) substrates using high-resolution transmission electron microscope (HRTEM), and indicated the rhombohedral HZO films to belong to the R3 space group [12].

The R-phase ferroelectric hafnia-based films with good crystal quality were reported to maintain a large polarization of 34 $\mu C/cm^2$ (comparable to the value of conventional ferroelectric $BaTiO_3$ [13,14]) and to not need wake-up cycling [11]. Compared to the well-known polar orthorhombic phase [15-22], the ferroelectric mechanisms and the structural properties of R-phase $HfO_2$ remain to be fully clarified. A comprehensive theoretical study is thus in demand, in order to understand some of the inconsistent experimental results and to reveal the impact of compressive in-plane lattice strain induced by substrates on the relative stability and polarization of R3 and R3m phases of HfO2.

In this work, we present a comparison of the relative stability and ferroelectricity in bulk R3 and R3m phases of HfO2. From the energetic perspective, they have similar total energies, when considering the equilibrium structures. The R3m phase possesses only a small polarization of 0.04 $\mu$C/cm$^2$, while the R3 phase with additional distortions presents a large polarization value of 41.14 $\mu$C/cm$^2$. We confirmed that the relative stability and ferroelectricity of R3m can be effectively enhanced by an in-plane compressive strain. Under a large compressive strain, the R3 phase will spontaneously transform to the R3m phase. On the other hand, the polarization direction and amplitude of the two R phases can be tuned by the in-plane strain, while the values of the ferroelectric energy barriers remain in a moderate range. To understand the nature of their ferroelectricity, we performed a symmetry mode analysis and found that the R phases are improper ferroelectrics with cooperative coupling between the ferroelectric mode and other distortion modes. These results highlight the interplay of structural stability and ferroelectricity upon compressive strain, which may help in understanding the experimental results of the hafnia R phases.

## II. COMPUTATIONAL DETAILS

Density functional theory (DFT) calculations were performed using the Vienna Ab initio simulation package (VASP) [23,24] with the projector-augmented wave (PAW) method [25]. The exchange-correlation functional follows the generalized gradient approximation (GGA) in the form of Perdew-Burke-Ernzerhof (PBE) [26]. The energy cutoff of plane wave basis was adopted to be 600 eV. The energy difference and force convergence criteria were set to be smaller than $10^{-7}$ eV and 0.005 eV/Å, respectively. The 5×5×4 Γ-centered Monkhorst-Pack *k*-grids were used for the calculations [27].

## III. RESULTS AND DISCUSSION

## A. Crystal structure and relative stability

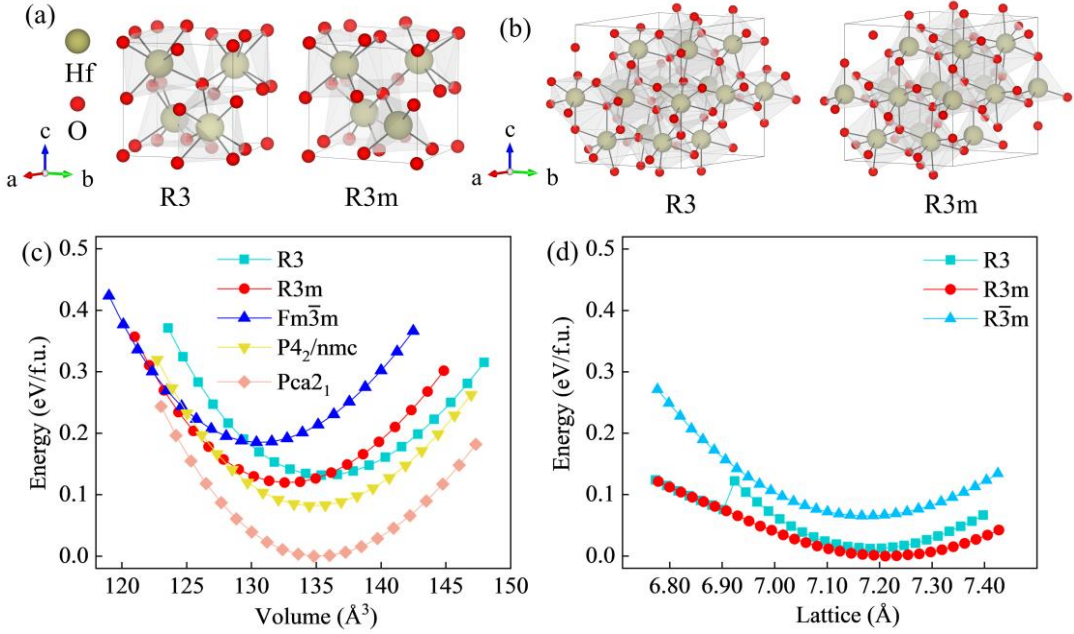

Figure 1. Atomic structure of (a) primitive cells and (b) conventional unit cells of R3 and R3m phase HfO$_2$. (c) Total energies versus volume of various phases of HfO$_2$. The zero of the energy scale is set as the energy of the Pca21 phase. (d) Total energies versus lattice constant *a* (under in-plane biaxial strain) of the unit cells in the R3, R3m and R$\bar{3}$m phases of HfO$_2$. The zero of the energy scale is set as the energy of the R3m phase.

Figure 1(a) displays the atomic structure of the primitive cells of R3 and R3m phases of HfO$_2$, each containing 4 Hf and 8 O atoms. In the equilibrium structure of the R3 phase, 1 Hf atom is bonded to 7 adjacent O atoms. In the R3m phase, three quarters of Hf atoms are bonded to 8 adjacent O atoms, and the remaining quarter of the Hf atoms are bonded to 4 adjacent O atoms. Figure 1(b) displays their standard unit cell structures, each containing 12 Hf and 24 O atoms, the *c* axis of which corresponds to the [111] direction of the primitive cell, as depicted in Figure 1(a). The optimized lattice constants are listed in Table I, (*a*/*c*) = 7.20/9.09 Å and 7.20/8.83 Å for R3 and R3m, respectively. They are in good agreement with previous theoretical studies [28,29], and both structures are non-centrosymmetric. Their polar point groups of 3 and 3m allow, from the symmetry point of view, for a macroscopic polarization, that will be discussed in detail below.

Table I. Lattice constants - $a$ and $c$ (Å) - and ferroelectric polarizations $P$ ($\mu$C/cm$^2$) of R3, R3m and R$\bar{3}$m phases, $\alpha$ denotes the angle of the primitive cell.

| Phase | $a$ (Å) | $c$ (Å) | $\alpha$ (°) | $P$ ($\mu$C/cm$^2$) |
|---|---|---|---|---|
| R3 | 7.20 | 9.09 | 88.81 | 41.14 |
| R3m | 7.20 | 8.83 | 88.72 | 0.04 |
| R$\bar{3}$m | 7.20 | 8.72 | 90.46 | 0.00 |

From the perspective of total energy, we show their energy-volume curves in Figure 1(d) and compare the relative stability with respect to other previously proposed structural phases (cfr Figure 1(c)). The cubic (Fm$\bar{3}$m) is the high temperature phase observed when the temperature exceeds 2773 K [30]; the cubic phase firstly changes to a tetragonal phase (P4$_2$/nmc) at 2773 K and then at 1973 K to a monoclinic (P2$_1$/c) phase [31,32], the latter being the HfO$_2$ most stable room temperature phase. Besides, the orthorhombic phase (Pca2$_1$) is the most studied phase of hafnia ferroelectric thin films [20,33-35]. The relative energies of R phases at their equilibrium volumes are noticeably higher than other phases, indicating the difficulty of a direct synthesis of R phases in HfO$_2$. However, the relative energy differences between all the phases are gradually converging as the volume decreases, implying the possibility of realization under high pressure conditions or on certain substrates with large compressive strain.

Indeed, R-phase HfO$_2$ films were reported to be successfully grown on substrates such as GaN(0001)/Si(111) [36] and LSMO/STO(001) [11,37] in the form of 50% Zr doping, where the in-plane strain should have a dominating impact. In Figure 1(d), we compared the relative stability of the three R phases by varying the in-plane biaxial strain. Firstly, we confirm that the R3m phase is always energetically more stable than the R3 phase under any strain. Furthermore, we found that the number of bonds of each Hf atom to adjacent O atoms in R3m phase is changed from the original coexistence of 4-coordination and 8-coordination to a 7-coordination structure when the lattice constant $a$ is smaller than 7.06 Å. When the in-plane lattice constant is smaller than 6.95 Å, there is a structural transition of R3 to the R3m phase. This collapse is confirmed by the analysis of the involved symmetry modes, lattice constants and bond length distribution (see the detailed comparison in Figure 2(a), Table SI and Table SII). As the overall relative energy difference between two ferroelectric R phases under in-plane strain is indeed small, experimentally they may show mixed coexisting forms [11].

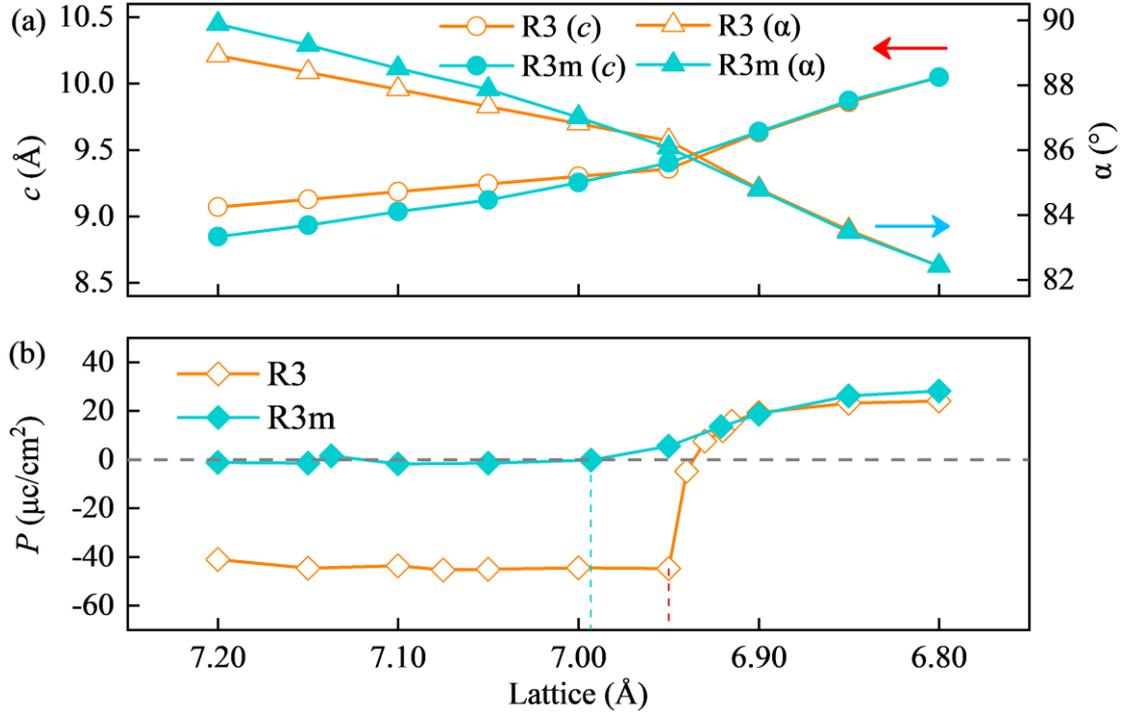

Figure 2. (a) Lattice constant $c$ (Å), R phase angle $\alpha$ (°), and (b) polarization of R3 and R3m phases as a function of in-plane lattice constant.

### B. Ferroelectricity under in-plane strain

As mentioned in previous studies [11,36,38], both the R3 and R3m-phases of $HfO_2$ are ferroelectric. In their equilibrium structures, the R3 phase shows a ferroelectric polarization of 41.14 $\mu C/cm^2$, which is comparable to 20-50 $\mu C/cm^2$ of the $Pca2_1$ phase [9,17,39-42] or that of the traditional ferroelectric perovskite $BaTiO_3$. On the other hand, the polarization of R3m phase of its equilibrium structure is two orders of magnitude smaller, i.e. only 0.04 $\mu C/cm^2$. The R3 phase intrinsically presents a much larger polarization than the R3m phase, due to the additional symmetry lowering. These systems present a strong polarization dependence with respect to in-plane strain. When the lattice constants of the R3m and R3 phases exceed 7.00 A and 6.95 A, respectively, the berry phase method indicates an abrupt increase in the polarization value of the structure, as depicted in Figure 2(b). We suggest that the changes in the polarization values of the R3 and R3m phases may be due to the rearrangement of electrons and related changes in the bond lengths, caused by the compressive strain (cfr Figure S2).

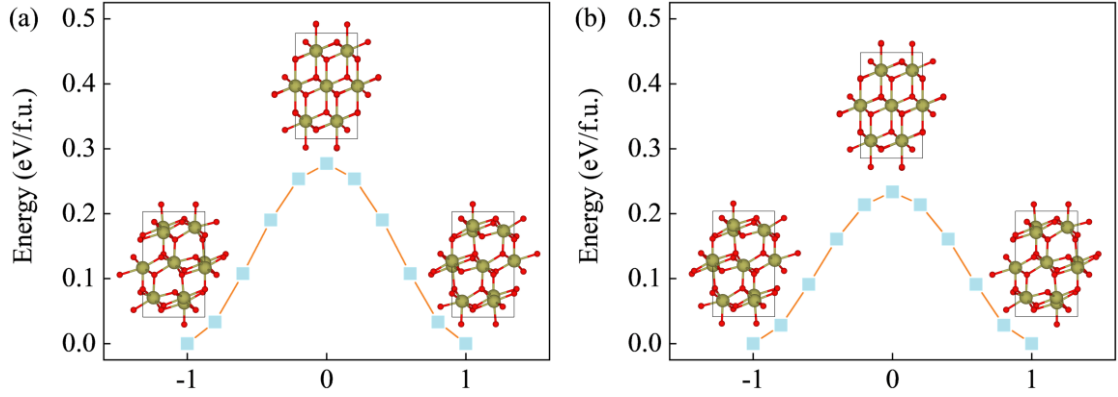

Figure 3. The ferroelectric switching pathway with (a) equilibrium lattice constants ($a = b = 7.20$ Å) of the R3 phase and (b) compressively strained ($a = b = 6.80$ Å) R3m phase. Here, 1, 0, and -1 correspond to the FE, PE and -FE states, respectively.

Figure 3 shows the energy barriers of the R3 and R3m phases during the ferroelectric switching. At the equilibrium lattice constant ($a = b = 7.20$ Å), the energy barrier of R3 phase is around 0.28 eV/f.u., while the energy barrier of R3m is around 0.10 eV/f.u, as seen in Figure S3(b). Under a large compressive strain ($a = b = 6.80$ Å), the energy barrier of the R3m phase is increased to ~0.23 eV/f.u., as shown in Figure 3(b). The energy barriers of the R3 phase and the R3m phase are equal, as shown in Figure S3(a). Overall, these values of energy barriers are larger than those of the orthorhombic ferroelectric phases, but still within a moderate range, indicating that the polarization of R-phase $HfO_2$ is likely to be switchable under an external electric field.

C. **Microscopic mechanisms of ferroelectricity**

To gain more insights on the different behavior of the polarization between the R3 and R3m phases, we performed a symmetry mode analysis by decomposing the distortion modes, according to the symmetry reduced from the high-symmetry structure ($R\bar{3}m$) to low-symmetry FE structures (R3 or R3m). We carried out the analyses for two cases: (i) R3 phase at its equilibrium lattice constant ($a = b = 7.20$ Å); (ii) R3m phase under compressive strain ($a = b = 6.80$ Å). The former is used to understand the origin of the large polarization of R3 in its equilibrium state, whereas the latter is of relevance for the polarization of the R-phase $HfO_2$ under a large compressive strain [43,44]. The mode analyses of the R3 phase under a compressive strain ($a = b = 6.80$ Å) and the R3m phase at its equilibrium lattice constant ($a = b = 7.20$ Å) are also discussed.

The mode decomposition of the ferroelectric R phases is based on the same high-symmetry reference phase with $R\bar{3}m$ space group. In order to simplify the description, we focus here only on the contribution of atomic displacements, thereby the volume and shape of the cell are fixed during the analysis [45]. The distortion modes of R3 and R3m phases are shown in Figure S4 and Figure S5, respectively. The corresponding amplitudes of these modes under different in-plane strains are shown in Table SIII. We found that all six distortion modes of the R3 phase individually increase the total energy. Among them, only the $F_2^-$ mode will spontaneously lower the crystal symmetry of $R\bar{3}m$. However, when the dimensionless amplitude $Q$ reaches 1, the $F_2^-$ mode itself still increases the total energy due to the excessive magnitude of distortions, as shown in Figure 4(a). In addition, the $F_2^-$ mode itself doesn't play an important role in the polarization value. It is the $\Gamma_2^-$ mode that contributes the most to the polarization value of 55.50 $\mu C/cm^2$, as listed in Table SIV. Thereby, we studied the multi-mode coupling to search for a softening of the $\Gamma_2^-$ mode. We found that a two-body interaction involving the $\Gamma_2^-$ mode is not enough to achieve softening, even by considering the coupling with the $F_2^-$ mode. The ferroelectric $\Gamma_2^-$ mode needs to couple with all the other five modes to decrease the total energy, as shown in Figure 4(b), indicating the improper nature of the polarization of the R3 phase. Here, the other five modes together actually increase the total energy by 0.17 eV/f.u. at first. After including the $\Gamma_2^-$ mode, an overall energy decrease of 0.22 eV/f.u. is achieved. Hence, the $\Gamma_2^-$ mode is essential in the stabilization of the R3 phase.

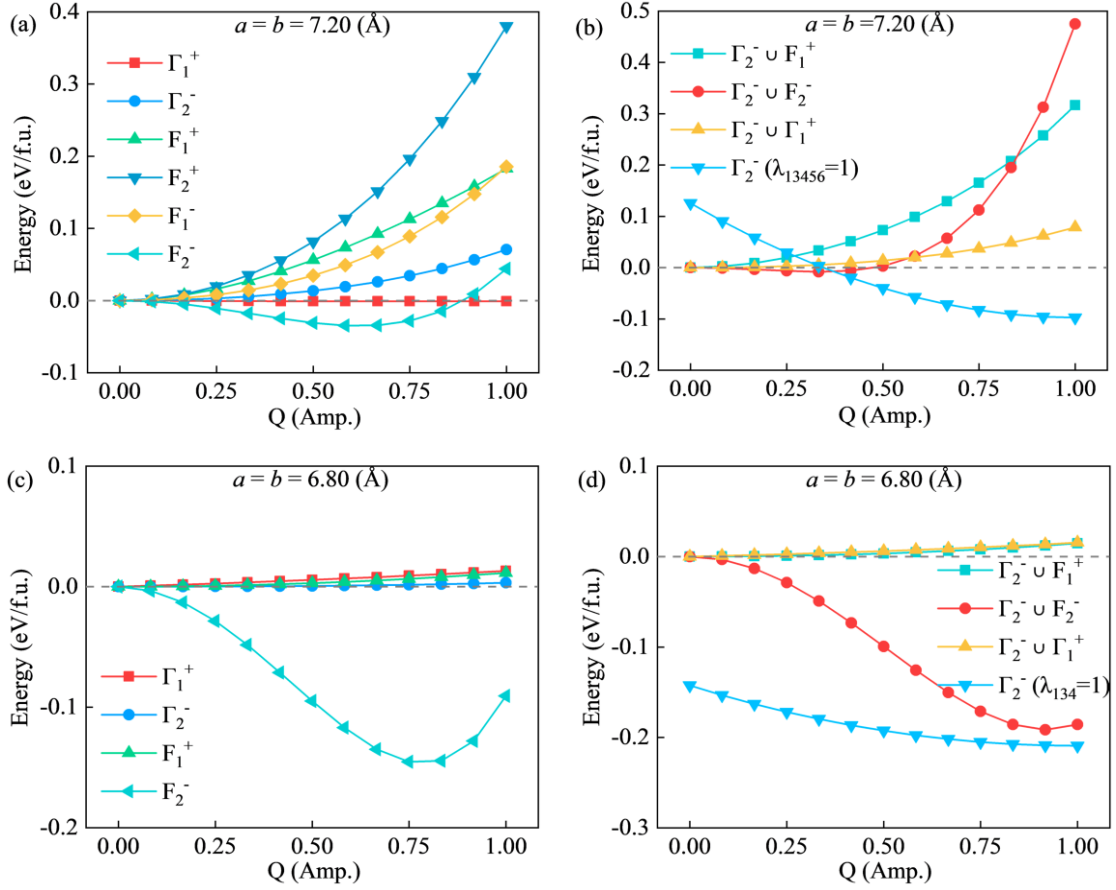

Figure 4. Total energies as a function of (a) symmetry modes of R3 phase at its equilibrium lattice constant ($a = b = 7.20$ Å), (b) Total energies as a function of combined symmetry modes; $\lambda_{13456}=1$ means $Q(\Gamma_1^+) = 1$, $Q(F_1^+) = 1$, $Q(F_2^+) = 1$, $Q(F_1^-) = 1$ and $Q(F_2^-) = 1$. (c) Total energies as a function of each symmetry mode of R3m phase under compressive strain ($a = b = 6.80$ Å). (d) Total energies as a function of each symmetry modes; $\lambda_{134}=1$ means $Q(\Gamma_1^+) = 1$, $Q(F_1^+) = 1$ and $Q(F_2^-) = 1$.

On the other hand, there are only four distortion modes of the R3m phase. Unlike that of R3 phase, the $F_2^-$ mode itself now can reduce the total energy by 0.09 eV/f.u., while the other modes still increase the total energy, as shown in Figure 4(c). Due to the dominating role of the $F_2^-$ mode in the decrease of total energy, the ferroelectric $\Gamma_2^-$ mode (that contributes a polarization of 26.40 $\mu$C/cm$^2$) can be softened in a two-body interaction with the $F_2^-$ mode with an energy decrease of 0.18 eV/f.u.. Finally, including four modes at the same time can altogether reduce the total energy by 0.21

eV/f.u., as shown in Figure 4(d). This indicates that the ferroelectricity of the R3m phase is also improper, and that the $F_2^-$ mode plays a dominating role in the stabilization of the $\Gamma_2^-$ mode. Upon increasing compressive strain, the amplitude of the ferroelectric mode $\Gamma_2^-$ of R3m phase is effectively increased from 0.00 Å to 0.18 Å, while the amplitude of $F_2^-$ mode is increased from 0.62 Å to 0.81 Å. Thereby, the coupling between these two modes acts cooperatively under compressive strain.

## IV. CONCLUSION

In summary, we performed a comparative theoretical study of the ferroelectric R3 and R3m phases of $HfO_2$. From the stability point of view, the two phases have a comparable total energy when the compressive strain is small. However, under a large compressive strain, the R3 phase will spontaneously transform to the R3m phase. Both the direction and amplitude of their polarization can be tuned by in-plane strain, while preserving the switch ability with moderate energy barriers. By means of symmetry mode analysis, the ferroelectricity is proved to be improper for the R phases. The cooperative couplings between the ferroelectric mode and other distortion modes are confirmed, and effectively controlled by strain. Our results could help understanding some key findings of the ferroelectric observations in doped hafnia thin films.

**ACKNOWLEDGMENT**

This work was supported by National Natural Science Foundation of China (12074241, 11929401, 52120204), Science and Technology Commission of Shanghai Municipality (22XD1400900, 20501130600, 21JC1402700, 21JC1402600, 22YF1413300), China Postdoctoral Science Foundation (2022M722035), Key Research Project of Zhejiang Laboratory (2021PE0AC02), High-Performance Computing Center, Shanghai Technical Service Center of Science and Engineering Computing, Shanghai University. S.P. acknowledges support from the Italian Ministry of Research under the PRIN-MUR project "TWEET: Towards Ferroelectricity in two dimensions" (IT-MIUR Grant No. 2017YCTB59).

**REFERENCE**

[1] C. Bowen, H. Kim, P. Weaver, and S. Dunn, Energy & Environmental Science **7**, 25 (2014).

[2] J. Scott, Science **315**, 954 (2007).

[3] R. Ramesh, Nature Materials **9**, 380 (2010).

[4] C. B. Eom and S. Trolier-McKinstry, Mrs Bulletin **37**, 1007 (2012).

[5] S. Mueller, S. R. Summerfelt, J. Muller, U. Schroeder, and T. Mikolajick, IEEE Electron Device Letters **33**, 1300 (2012).

[6] A. Chernikova, M. Kozodaev, A. Markeev, D. Negrov, M. Spiridonov, S. Zarubin, O. Bak, P. Buragohain, H. Lu, and E. Suvorova, ACS Applied Materials & Interfaces **8**, 7232 (2016).

[7] J. Müller, P. Polakowski, S. Mueller, and T. Mikolajick, ECS Journal of Solid State Science and Technology **4**, N30 (2015).

[8] U. Schroeder, S. Mueller, J. Mueller, E. Yurchuk, D. Martin, C. Adelmann, T. Schloesser, R. van Bentum, and T. Mikolajick, ECS Journal of Solid State Science and Technology **2**, N69 (2013).

[9] T. Böscke, J. Müller, D. Bräuhaus, U. Schröder, and U. Böttger, Applied Physics Letters **99**, 102903 (2011).

[10] X. Sang, E. D. Grimley, T. Schenk, U. Schroeder, and J. M. LeBeau, Applied Physics Letters **106** (2015).

[11] Y. Wei, P. Nukala, M. Salverda, S. Matzen, H. J. Zhao, J. Momand, A. S. Everhardt, G. Agnus, G. R. Blake, and P. Lecoeur, Nature Materials **17**, 1095 (2018).

[12] L. Bégon-Lours, M. Mulder, P. Nukala, S. De Graaf, Y. A. Birkhölzer, B. Kooi, B. Noheda, G. Koster, and G. Rijnders, Physical Review Materials **4**, 043401 (2020).

[13] N. Iles, K. D. Khodja, A. Kellou, and P. Aubert, Computational Materials Science **87**, 123 (2014).

[14] E. Goh, L. Ong, T. Yoon, and K. Chew, Computational Materials Science **117**, 306 (2016).

[15] E. Yurchuk, J. Müller, S. Knebel, J. Sundqvist, A. P. Graham, T. Melde, U. Schröder, and T. Mikolajick, Thin Solid Films **533**, 88 (2013).

[16] M. H. Park, Y. H. Lee, H. J. Kim, Y. J. Kim, T. Moon, K. D. Kim, J. Mueller, A. Kersch, U. Schroeder, and T. Mikolajick, Advanced Materials **27**, 1811 (2015).

[17] S. Mueller, J. Mueller, A. Singh, S. Riedel, J. Sundqvist, U. Schroeder, and T. Mikolajick, Advanced Functional Materials **22**, 2412 (2012).

[18] U. Schroeder, E. Yurchuk, J. Müller, D. Martin, T. Schenk, P. Polakowski, C. Adelmann, M. I. Popovici, S. V. Kalinin, and T. Mikolajick, Japanese Journal of Applied Physics **53**, 08LE02 (2014).

[19] J. Müller, T. Böscke, D. Bräuhaus, U. Schröder, U. Böttger, J. Sundqvist, P. Kücher, T. Mikolajick, and L. Frey, Applied Physics Letters **99**, 112901 (2011).

[20] M. Hoffmann, U. Schroeder, T. Schenk, T. Shimizu, H. Funakubo, O. Sakata, D. Pohl, M. Drescher, C. Adelmann, and R. Materlik, Journal of Applied Physics **118**, 072006 (2015).

[21] T. Mittmann, M. Michailow, P. D. Lomenzo, J. Gärtner, M. Falkowski, A. Kersch, T. Mikolajick, and U. Schroeder, Nanoscale **13**, 912 (2021).

[22] M. Materano, T. Mittmann, P. D. Lomenzo, C. Zhou, J. L. Jones, M. Falkowski, A. Kersch, T. Mikolajick, and U. Schroeder, ACS Applied Electronic Materials **2**, 3618 (2020).

[23] P. Hohenberg and W. Kohn, Physical Review **136**, B864 (1964).

[24] B. Tong and L. Sham, Physical Review **144**, 1 (1966).

[25] P. E. Blöchl, Physical Review B **50**, 17953 (1994).

[26] J. P. Perdew, K. Burke, and M. Ernzerhof, Physical Review Letters **77**, 3865 (1996).

[27] H. J. Monkhorst and J. D. Pack, Physical Review B **13**, 5188 (1976).

[28] Y. Zhang, Q. Yang, L. Tao, E. Y. Tsymbal, and V. Alexandrov, Physical Review Applied **14**, 014068 (2020).

[29] F. Delodovici, P. Barone, and S. Picozzi, Physical Review B **106**, 115438 (2022).

[30] T. D. Huan, V. Sharma, G. A. Rossetti Jr, and R. Ramprasad, Physical Review B **90**, 064111 (2014).

[31] C. Curtis, L. Doney, and J. Johnson, Journal of the American Ceramic Society **37**, 458 (1954).

[32] D. Zhao, Z. Chen, and X. Liao, Microstructures **2**, 2022007 (2022).

[33] X. Sang, E. D. Grimley, T. Schenk, U. Schroeder, and J. M. LeBeau, Applied Physics Letters **106**, 162905 (2015).

[34] S. Clima, D. Wouters, C. Adelmann, T. Schenk, U. Schroeder, M. Jurczak, and G. Pourtois, Applied Physics Letters **104**, 092906 (2014).

[35] J. Müller, U. Schröder, T. Böscke, I. Müller, U. Böttger, L. Wilde, J. Sundqvist, M. Lemberger, P. Kücher, and T. Mikolajick, Journal of Applied Physics **110**, 114113 (2011).

[36] L. Bégon-Lours, M. Mulder, P. Nukala, S. de Graaf, Y. A. Birkhölzer, B. Kooi, B. Noheda, G. Koster, and G. Rijnders, Physical Review Materials **4** (2020).

[37] P. Nukala, Y. Wei, V. de Haas, Q. Guo, J. Antoja-Lleonart, and B. Noheda, Ferroelectrics **569**, 148 (2020).

[38] Y. Zhang, Q. Yang, L. Tao, E. Y. Tsymbal, and V. Alexandrov, Physical Review Applied **14**, 014068 (2020).

[39] T. Böscke, S. Teichert, D. Bräuhaus, J. Müller, U. Schröder, U. Böttger, and T. Mikolajick, Applied Physics Letters **99**, 112904 (2011).

[40] J. Muller, T. S. Boscke, U. Schroder, S. Mueller, D. Brauhaus, U. Bottger, L. Frey, and T. Mikolajick, Nano Letters **12**, 4318 (2012).

[41] S. Mueller, C. Adelmann, A. Singh, S. Van Elshocht, U. Schroeder, and T. Mikolajick, ECS Journal of Solid State Science and Technology **1**, N123 (2012).

[42] P. D. Lomenzo, P. Zhao, Q. Takmeel, S. Moghaddam, T. Nishida, M. Nelson, C. M. Fancher, E. D. Grimley, X. Sang, and J. M. LeBeau, Journal of Vacuum Science & Technology B, Nanotechnology and Microelectronics: Materials, Processing, Measurement, and Phenomena **32**, 03D123 (2014).

[43] Y. Zhang, Q. Yang, L. Tao, E. Y. Tsymbal, and V. Alexandrov, Physical Review Applied **14** (2020).

[44]  F. Delodovici, P. Barone, and S. Picozzi, Physical Review B **106**, 115438 (2022).
[45]   Y. Yang, F. Lou, and H. Xiang, Nano Letters **21**, 3170 (2021).

# Supplementary materials

**Structural stability and polarization analysis of rhombohedral phases of $HfO_2$**

Wenbin Ouyang[1], Fanhao Jia[1], Silvia Picozzi [2] and Wei Ren[1,*]

[1]*Physics Department, International Center for Quantum and Molecular Structures, Materials Genome Institute, State Key Laboratory of Advanced Special Steel, Shanghai Key Laboratory of High Temperature Superconductors, Shanghai University, Shanghai 200444, China*

[2] *Consiglio Nazionale delle Ricerche CNR-SPIN, c/o Univ. "G. D'Annunzio", 66100 Chieti, Italy*

\* *renwei@shu.edu.cn*

Table SI. The in-plane lattice parameters *a* and *b* of R3, R3m and R$\bar{3}$m are fixed, and the optimized lattice constants along the *c*-direction are given.

| $a = b$ | 6.80 (Å) | 6.90 (Å) | 7.00 (Å) |
|---|---|---|---|
| *c* (R3) | 10.05 | 9.63 | 9.30 |
| *c* (R3m) | 10.05 | 9.63 | 9.25 |
| *c* (R$\bar{3}$m) | 9.44 | 9.24 | 9.05 |

Table SII. Statistics of bond lengths (in Å) and numbers of Hf-O bonding when the in-plane lattice parameters *a* and *b* of R3 and R3m are fixed at 6.80 Å, 6.90 Å, and 7.00 Å.

| | | Bond Length | | | | | | | | |
|---|---|---|---|---|---|---|---|---|---|---|
| Lattice | Phase | 2.02 | 2.05 | 2.12 | 2.16 | 2.18 | 2.22 | 2.28 | 2.84 | 3.11 |
| 6.8 Å | R3 | 7 | 9 | 10 | 5 | 6 | 10 | 5 | 6 | 2 |
| | R3m | 7 | 9 | 10 | 5 | 6 | 10 | 5 | 6 | 2 |
| | | 2.02 | 2.05 | 2.12 | 2.16 | 2.18 | 2.22 | 2.28 | 2.84 | 3.11 |
| 6.9 Å | R3 | 7 | 2 | 7 | 16 | 10 | 5 | 5 | 6 | 2 |
| | R3m | 7 | 2 | 7 | 16 | 10 | 5 | 5 | 6 | 2 |
| | | 2.04 | 2.07 | 2.11 | 2.12 | 2.14 | 2.15 | 2.19 | 2.30 | 2.92 |
| 7.0 Å | R3 | 6 | 5 | 9 | 3 | 1 | 7 | 9 | 7 | 3 |
| | | 2.02 | 2.03 | 2.06 | 2.15 | 2.17 | 2.22 | 2.24 | 2.33 | 2.39 |
| 7.0 Å | R3m | 7 | 2 | 7 | 10 | 6 | 5 | 10 | 5 | 6 |

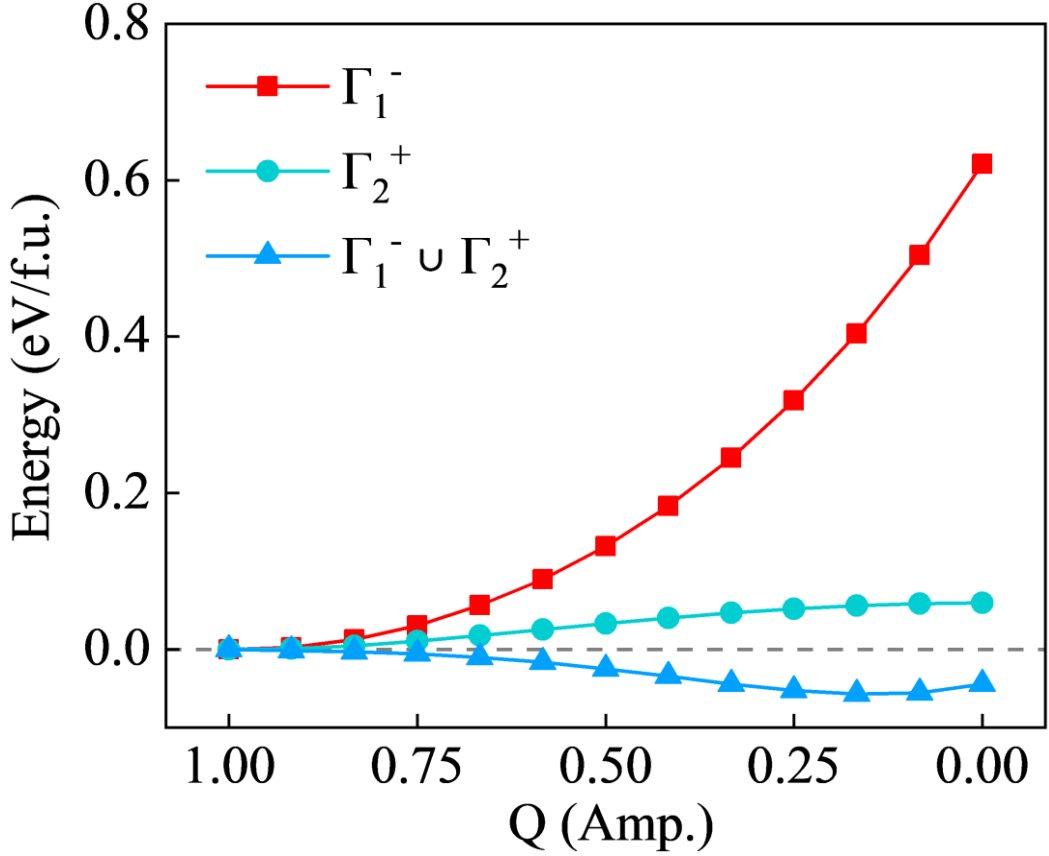

Figure S1. Total energy as a function of symmetry modes $\Gamma_1^-$ and $\Gamma_2^+$ of $HfO_2$ under fixed lattice constants ($a = b$ = 6.91 Å and $c$ = 9.51 Å).

To confirm the structural transition from R3 to R3m under large compressive in-plane strain, we considered three in-plane lattice constants, namely 6.80 Å, 6.90 Å, and 7.00 Å, respectively. We first compared the lattice constant $c$ of 'R3' and R3m in Table SI. It is easy to find that the $c$ of R3 is noticeably larger than R3m when $a$ = 7.00 Å, while they are nearly identical to each other when $a$ is 6.90 or 6.80 Å. Then, we further compared their bond length distributions in Table SII. When $a$ is 6.90 or 6.80 Å, the bond lengths are matched one-by-one between the 'R3' and R3m calculations, which strongly confirms that they are indeed the same structure. To understand how the structural transition happens, we performed a symmetry mode analysis. We found that an in-phase distortion mode $\Gamma_1^-$ of R3m combined with a $\Gamma_2^+$ mode that only contains in-plane distortions are responsible for this transition. The cooperative coupling of these two modes softens the R3 structure, resulting in a spontaneous energy lowering to the

R3m symmetry.

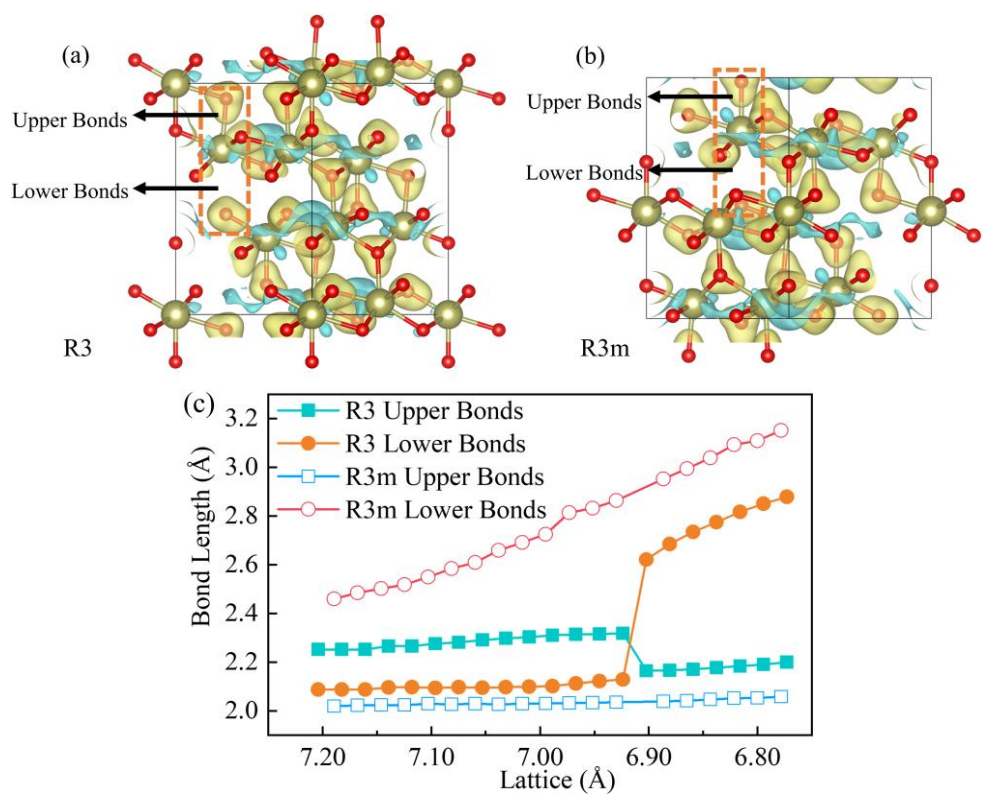

Figure S2. Schematic diagram of the differential charge density in (a) the R3 and (b) the R3m phase equilibrium states and the Hf-O bond lengths. (c) The variation of Hf-O bond lengths in R3 and R3m phases with different in-plane lattice parameters.

We present two Hf-O bond lengths according to the bonding between the hafnium atom and the upper / lower oxygen atoms, namely upper and lower Hf-O bonds, respectively, as shown in the orange dashed boxes in Fig. S2(a) and Fig. S2(b). Figure S2(c) gives the variation pattern of Hf-O bond lengths for R3 and R3m phases at different lattice parameters. The polarization curves tend to be constants for both R3 and R3m phases when the lattice parameter is in the range of 7.20 Å to 6.95 Å. The Hf-O bond length in the upper layer of R3m phase remains at 2.05 Å, while the Hf-O bond length in the lower layer increases from 2.45 Å to 2.70 Å. The polarization value of R3m phase remains around 0 μC/cm² during this interval. The R3 upper Hf-O bond length is around 2.25 Å, the lower Hf-O bond length is around 2.10 Å, and the polarization value stays around -40 μC/cm2. When the lattice parameter is further reduced to below 6.95 Å, the Hf-O bond lengths of the upper and lower layers of the R3 phase show large changes and tend to be closer to the R3m phase, which corresponds to the phase transition from the R3 phase to the R3m phase. For the R3m phase, the upper Hf-O bond length still keeps a value around 2.05 Å, but the lower Hf-O bond length increases further with the compressive strain, eventually leading to a significant increase in the polarization value of the R3m phase. This can be interpreted in terms of an increase in the distance between the centers of mass of positive and negative charges due to the increase in the bond length, i.e. to an increase in the value of the electric dipole moment per unit volume.

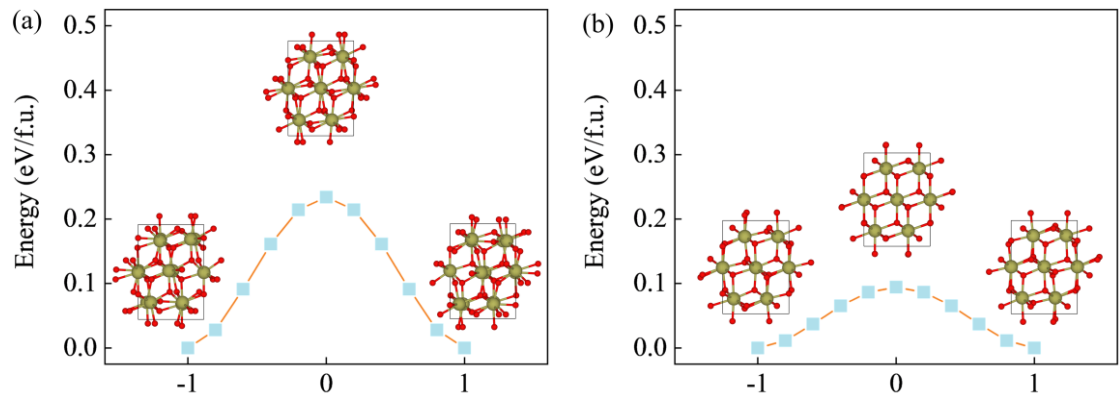

Figure S3. The ferroelectric switching pathway of the R3 phase and R3m phase with an in-plane lattice parameter of (a) 6.80 Å and (b) 7.20 Å. Here, 1, 0, and -1 correspond to the FE, PE and -FE states, respectively.

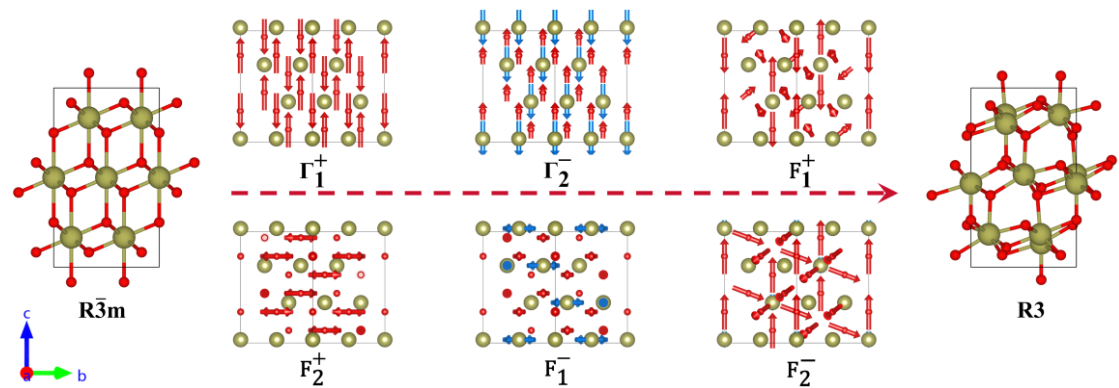

Figure S4. The atomic displacements of each single mode in the R3 phase, using $R\bar{3}m$ as the high-symmetry reference phase.

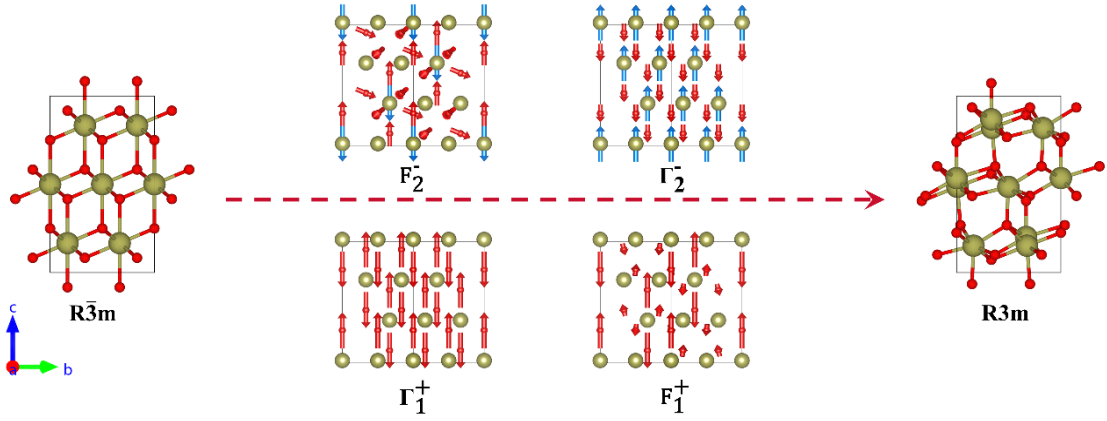

Figure S5. The atomic displacements of each single mode in the R3m phase, using R$\bar{3}$m as the high-symmetry reference phase.

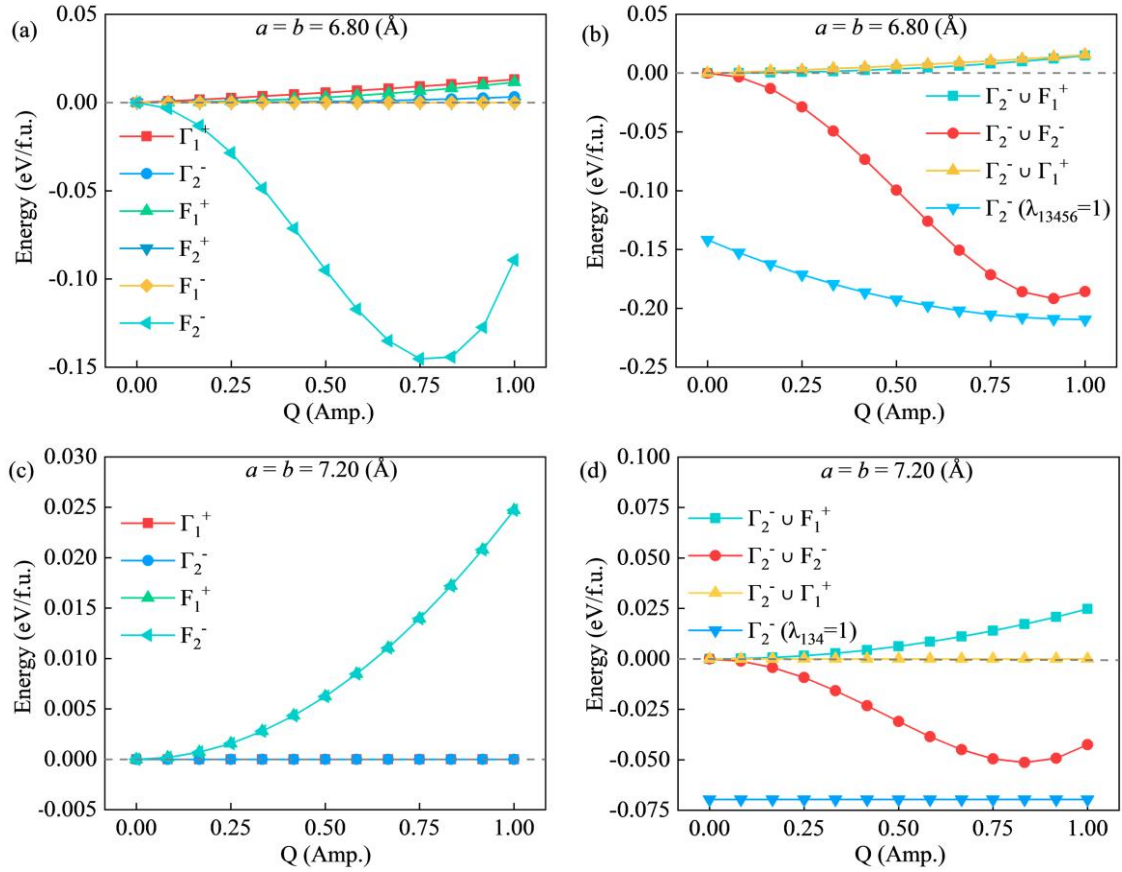

Figure S6. Total energies as function of symmetry modes. (a) R3 phase under compressive strain ($a = b = 6.80$ Å). (b) Total energies of each symmetry mode, $\lambda_{13456} = 1$ means $Q(\Gamma_1^+) = 1$, $Q(F_1^+) = 1$, $Q(F_2^+) = 1$, $Q(F_1^-) = 1$ and $Q(F_2^-) = 1$. (c) R3m phase with equilibrium lattice constant ($a = b = 7.20$ Å). (d) Total energies of each symmetry mode, $\lambda_{134}=1$ means $Q(\Gamma_1^+) = 1$, $Q(F_1^+) = 1$ and

$Q(F_2^-) = 1$.

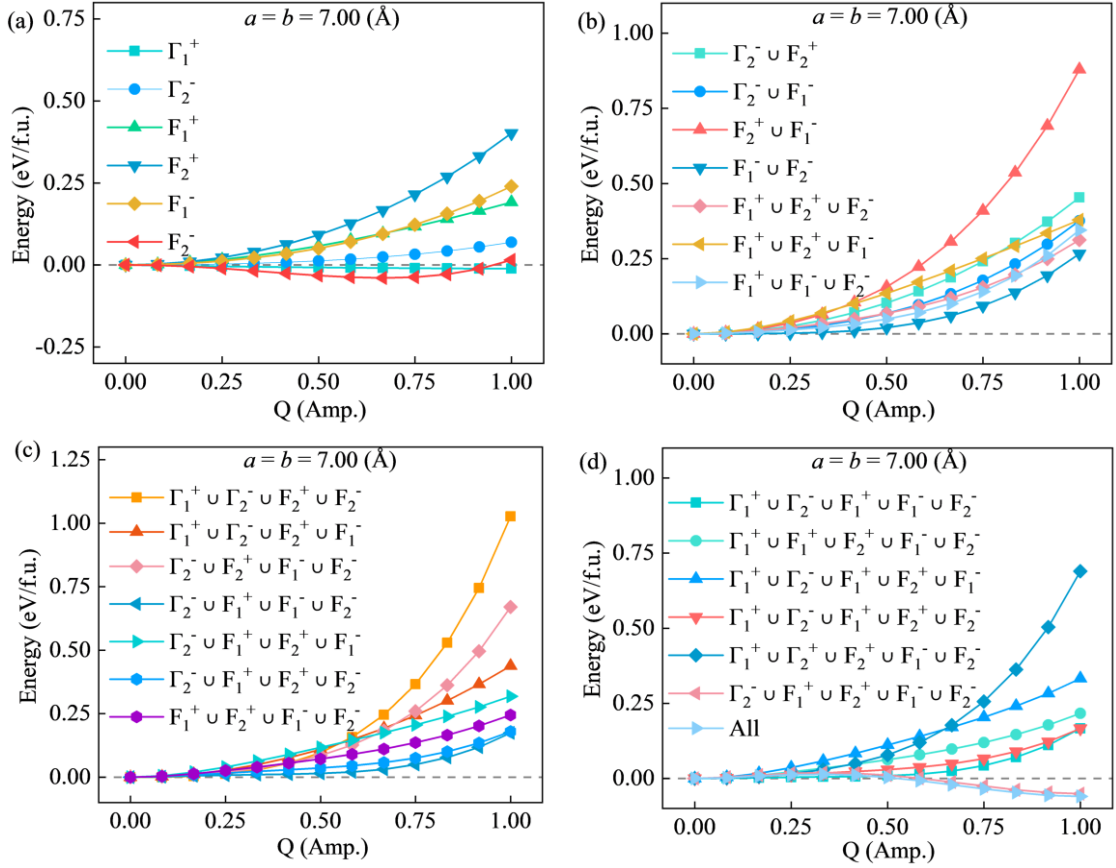

Figure S7. Atomic displacements of each single mode ordered by decreasing amplitude. $\Gamma_1^+$, $\Gamma_2^-$, $F_1^+$, $F_2^+$, $F_1^-$ and $F_2^-$ modes of R3 under the compressive strain ($a = b = 7.00$ Å). The panels report the *ab-initio* total energy for (a) single, (b) two coupled and three coupled modes (c) four coupled modes and (d) five coupled modes as a function of the generalized displacement coordinate Q.

Only one unstable soft mode, i.e. $F_2^-$, exists in the R3 phase, however the energy does not decrease further as more modes are coupled to the $F_2^-$ mode, as shown in Figure S7(b) and Figure S7(c). Only when almost all modes are coupled together can a ferroelectric distortion be induced in the $R\bar{3}m$ phase, as shown in Figure S7(d). Compared to the R3 phase, the R3m phase is easier to obtain both in terms of stability and in terms of the energy of the ferroelectric distortion pattern.

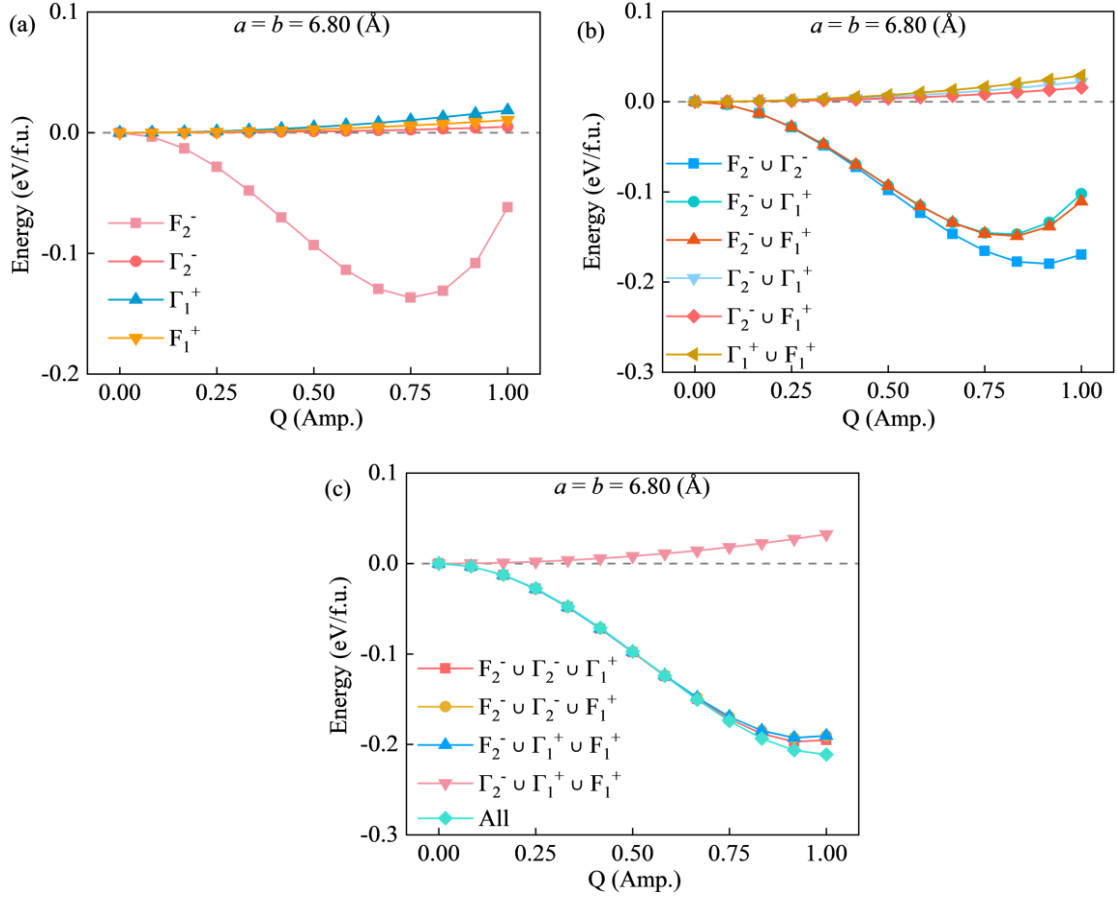

Figure S8. Atomic displacements of each single mode ordered by decreasing amplitude. $\Gamma_1^+$, $\Gamma_2^-$, $F_1^+$ and $F_2^-$ modes of R3m under the compressive strain ($a = b = 6.80$ Å). The panel reports the *ab*-initio total energy for (a) single, (b) two and (c) three coupled modes as a function of the generalized displacement coordinate Q.

The only soft mode is $F_2^-$, which has an instability of 0.14 eV/f.u at Q = 0.75. When the $F_2^-$ mode is coupled separately with $\Gamma_1^+$, $F_1^+$, $\Gamma_2^-$, respectively, there is an energy reduction of 0.14 eV/f.u, 0.15 eV/f.u and 0.18 eV/f.u at 0.75 < Q <1, as shown in Figure S8(b). When $F_2^-$ is coupled with $\Gamma_2^- \cup \Gamma_1^+$, $\Gamma_2^- \cup F_1^+$ and $\Gamma_1^+ \cup F_1^+$, the energy is reduced by 0.20 eV/f.u, 0.19 eV/f.u and 0.19 eV/f.u at Q = 1, as shown in Figure S8(c). Based on the above results, we can infer that the maximum allowed displacement Q is 0.75 only when the $F_2^-$ mode is alone; when we consider more modes coupled to the $F_2^-$ mode, the maximum allowed displacement Q of the mode increases while the energy becomes more stable, i.e. a typical behavior of an improper FE.

Table SIII. The magnitude of the effective displacement of the different modes of R3 and R3m.

| Lattice | Phase | Irrep | | | | | |
|---|---|---|---|---|---|---|---|
| | | $\Gamma_1^+$ | $\Gamma_2^-$ | $F_1^+$ | $F_2^+$ | $F_1^-$ | $F_2^-$ |
| 7.20 Å | R3 | 0.031 | 0.352 | 0.843 | 0.752 | 0.405 | 0.756 |
| | R3m | 0.001 | 0.001 | 0.087 | / | / | 0.616 |
| | | $\Gamma_1^+$ | $\Gamma_2^-$ | $F_1^+$ | $F_2^+$ | $F_1^-$ | $F_2^-$ |
| 7.10 Å | R3 | 0.043 | 0.377 | 0.854 | 0.725 | 0.398 | 0.746 |
| | R3m | 0.014 | 0.026 | 0.125 | / | / | 0.621 |
| | | $\Gamma_1^+$ | $\Gamma_2^-$ | $F_1^+$ | $F_2^+$ | $F_1^-$ | $F_2^-$ |
| 7.00 Å | R3 | 0.055 | 0.390 | 0.835 | 0.670 | 0.378 | 0.733 |
| | R3m | 0.020 | 0.021 | 0.168 | / | / | 0.636 |
| | | $\Gamma_1^+$ | $\Gamma_2^-$ | $F_1^+$ | $F_2^+$ | $F_1^-$ | $F_2^-$ |
| 6.90 Å | 'R3' | 0.008 | 0.100 | 0.115 | 0.001 | 0.001 | 0.719 |
| | R3m | 0.009 | 0.101 | 0.116 | / | / | 0.720 |
| | | $\Gamma_1^+$ | $\Gamma_2^-$ | $F_1^+$ | $F_2^+$ | $F_1^-$ | $F_2^-$ |
| 6.80 Å | 'R3' | 0.041 | 0.185 | 0.091 | 0.002 | 0.001 | 0.814 |
| | R3m | 0.041 | 0.185 | 0.091 | / | / | 0.812 |

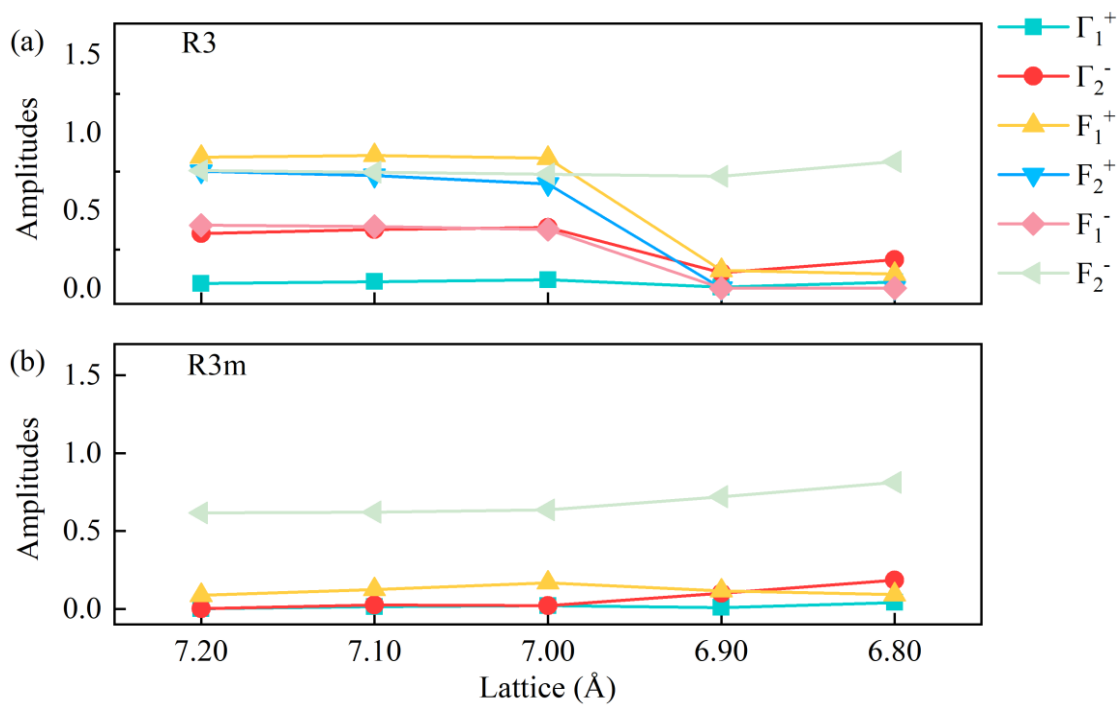

Figure S9. Relationship between lattice constants and mode weights in R3 and R3m phases

Table SIV. Born effective charge method is used to calculate the contribution of different modes of R3 and R3m to the polarization values in $\mu C/cm^2$.

| Lattice | Phase | Irrep | | | | | |
|---|---|---|---|---|---|---|---|
| | | $\Gamma_1^+$ | $\Gamma_2^-$ | $F_1^+$ | $F_2^+$ | $F_1^-$ | $F_2^-$ |
| 7.20 Å | R3 | 0.414 | -55.465 | -3.461 | 0 | 0 | 3.931 |
| | R3m | -0.004 | -0.167 | -0.113 | / | / | -0.054 |
| | | $\Gamma_1^+$ | $\Gamma_2^-$ | $F_1^+$ | $F_2^+$ | $F_1^-$ | $F_2^-$ |
| 7.10 Å | R3 | -0.001 | -58.149 | 0.002 | 0 | 0 | -0.953 |
| | R3m | -0.073 | 3.914 | -0.433 | / | / | -1.608 |
| | | $\Gamma_1^+$ | $\Gamma_2^-$ | $F_1^+$ | $F_2^+$ | $F_1^-$ | $F_2^-$ |
| 7.00 Å | R3 | 0 | -59.980 | -0.003 | 0 | 0 | -0.094 |
| | R3m | 0 | 3.228 | 0 | / | / | 0.012 |
| | | $\Gamma_1^+$ | $\Gamma_2^-$ | $F_1^+$ | $F_2^+$ | $F_1^-$ | $F_2^-$ |
| 6.90 Å | 'R3' | 0 | 14.957 | -0.003 | 0 | 0 | 0.006 |
| | R3m | 0 | 14.957 | -0.003 | / | / | 0.006 |
| | | $\Gamma_1^+$ | $\Gamma_2^-$ | $F_1^+$ | $F_2^+$ | $F_1^-$ | $F_2^-$ |
| 6.80 Å | 'R3' | -0.003 | 26.402 | -0.003 | 0 | 0 | -0.188 |
| | R3m | -0.003 | 26.402 | -0.003 | / | / | -0.188 |